\documentclass[english, reprint, aps, prb, twocolumn, superscriptaddress, showpacs, floatfix]{revtex4-1}
\usepackage{babel}
\usepackage{amssymb}
\usepackage{mathrsfs} 
\usepackage{amsmath}
\usepackage{float}
\usepackage{bbm}
\usepackage{graphicx}
\usepackage{epsfig}
\usepackage{verbatim}
\usepackage{bm}
\usepackage[usenames]{color}

\usepackage{hyperref}
\begin{document}

\title{Energy spectrum, exchange interaction and gate crosstalk in a pair of double-quantum-dot system: a molecular orbital calculation}

\author{Xu-Chen Yang}
\affiliation{Department of Physics and Materials Science, City University of Hong Kong, Tat Chee Avenue, Kowloon, Hong Kong SAR, China}
\affiliation{City University of Hong Kong Shenzhen Research Institute, Shenzhen, Guangdong 518057, China}
\author{Xin Wang}
\email{Corresponding author: x.wang@cityu.edu.hk}
\affiliation{Department of Physics and Materials Science, City University of Hong Kong, Tat Chee Avenue, Kowloon, Hong Kong SAR, China}
\affiliation{City University of Hong Kong Shenzhen Research Institute, Shenzhen, Guangdong 518057, China}
\date{\today}

\begin{abstract}
We present a theoretical study of a four-electron four-quantum-dot system based on molecular orbital methods, which hosts a pair of singlet-triplet spin qubits. We explicitly take into account of the admixture of electron wave functions in all dots, and have found that this mixing of wave functions has consequences on the energy spectrum, exchange interaction and the gate crosstalk of the system. Specifically, we have found that when the two singlet-triplet qubits are close enough, some of the states are no longer dominated by the computational basis states and the exchange interaction can not simply be understood as the energy difference between the singlet and triplet states. Using the Hund-Mulliken calculation of the Hubbard parameters, we characterize the effective exchange interaction of the system and have found good agreement with results calculated by taking energy differences where applicable. We have studied the two commonly conceived schemes coupling two qubits, the exchange and capacitive coupling, and have found that when the inter-qubit distance is at certain intermediate values, the two  kinds of coupling are comparable in strength, complicating analyses of the evolution of the two qubits. We also investigate the gate crosstalk in the system due to the quantum mechanical mixing of electron states and have found that while this effect is typically very weak, it should not be ignored if the spacing between the qubits are similar to or less than the distance between the double dots that constitute the qubit.
\end{abstract}


\maketitle

\section{introduction}

One of the key issues in the physical realization of quantum computing\cite{nielsen2010quantum} is the scalability, namely the ability of fabricating and coherently manipulating one or few qubits must be extended to an array of them. Spin qubits in semiconductor quantum dots serve as one of the most promising candidates for quantum computing, not only because of their long coherence times and high control fidelities,\cite{Petta.05, Bluhm.10b, Barthel.10, Maune.12, Pla.12, Pla.13, Muhonen.14, Kim.14,Kawakami.16} but also due to their prospect for scaling-up,\cite{Taylor.05} thanks to the mature infrastructure of present-day semiconductor industry. Various types of spin qubits are currently being pursued in laboratories worldwide, including the single-spin qubit being encoded in the spin up and down states of one electron,\cite{Loss.98,Muhonen.14} the singlet-triplet (S-T) qubit\cite{Petta.05, Maune.12} which employs two-electron singlet and triplet states as computational bases,\cite{Levy.02} the exchange-only qubit\cite{DiVincenzo.00,Laird.10} together with its variant, the resonant-exchange qubit,\cite{Medford.13} utilizing certain three-spin states in a triple-quantum-dot system, as well as the hybrid qubit\cite{Shi.12, Kim.14} which similarly uses three electrons but has them hosted by double quantum dots. Over the past decade, the S-T qubit has received intense attention because it is the simplest type of spin qubit which can be manipulated efficiently by pure electrical means. The vast success achieved at its single-qubit level\cite{Petta.05, Foletti.09,Bluhm.10c,Brunner.11,Maune.12, Petersen.13,Wu.14} has encouraged researchers to go on to two S-T qubits, which require four quantum dots occupied by four electrons.\cite{HansonBurkard.07, Trifunovic.12, Wardrop.14, Srinivasa.15}  For example, it has been shown that a single exchange pulse is sufficient to perform a two-qubit gate using a pair of exchange-coupled S-T qubits.\cite{Wardrop.14} Devices with four or even more quantum dots can now be fabricated in the laboratory,\cite{Zajac.16} and progresses have been made toward precise manipulation of two S-T qubits.\cite{vanWeperen.11,Shulman.12, Nichol.16}
Nevertheless, coherent control of the collective multi-electron states has remained challenging.

The microscopic theoretical studies of double quantum dots, in particular those from the molecular orbital theory,\cite{Burkard.99,Hu.00}
have played an important role in elucidating the physics of an S-T qubit. Not only the energy spectra and exchange interaction of the S-T qubit have been investigated in detail for their dependences on the dimensions of the device and the external fields,\cite{Burkard.99, Hu.00, He.05, Li.10, Mehl.14} the responses of the qubit to various types of noises are also extensively studied.\cite{Nielsen.10, Raith.11, Barnes.11, Bakker.15} However, the extensions of these studies to two S-T qubits involving four quantum dots are less well explored. Protocols to perform entangling gates have been put forward for capacitively coupled S-T qubits,\cite{Stepanenko.07, Saraiva.07, Ramon.11,Nielsen.12,Calderon.15,Srinivasa.15,Wang.15} and the ``sweet spots'' on the energy spectra where the control is maximally immune to noises have been suggested.\cite{Ramon.11,Yang.11b}   Alternative proposals making use of an additional quantum state which mediates the two-qubit coupling have also been suggested theoretically to be able to achieve high-fidelity entangling gates,\cite{Mehl.14b} which is currently being pursued in laboratories.
While the molecular-orbital-theoretic calculations similar to that of one S-T qubit have been carried out in many of these works, the two S-T qubits are always assumed to be quantum-mechanically ``well-separated'', that is, the wave function of one S-T qubit is assumed to have no admixture of that from the other.\cite{Yang.11b,Calderon.15} This has lead to a substantial reduction of the dimensionality of the Hilbert space and has greatly facilitated the calculations since the two S-T qubits are treated essentially separately and are only connected by a Coulomb integral.

In this work we conduct a comprehensive theoretical study of the four-quantum-dot system via the molecular orbital theory. In particular, we allow for the cases in which the admixture between wave functions of the two S-T qubits cannot be neglected. We find that the energy spectra, while being very similar to that of the simplified case in which the two S-T qubits are treated separately, show important features of mixing as the two S-T qubits become  close. We show that the qubit states may be less well defined for certain parameters, and consequently the exchange interaction that is usually defined as the energy difference between the singlet and triplet states must be understood as an effective exchange interaction that can be derived from the Hubbard parameters calculated microscopically.  Using the direct access to wave functions, we quantitatively calculate the strength of the capacitive and exchange coupling between the two S-T qubits and perform a comparison between them. We further study the gate crosstalk, the situation in which the manipulation of one qubit inevitably affects the other idle one due to the admixture of electron wave functions between the two qubits. We find that while the gate crosstalk via this quantum channel is typically weak, it can be much pronounced when the distance between the two qubits are small enough. 

The remainder of this paper is organized as follows. In Sec.~\ref{sec:model} we present the model and methods used in this work. We then present results in Sec.~\ref{sec:res}, including the energy spectrum (Sec.~\ref{sec:spectrum}), the exchange interaction (Sec.~\ref{sec:exch}) and the gate crosstalk (Sec.~\ref{sec:crosstalk}). We conclude in Sec.~\ref{sec:conclusion}.

\begin{figure}
  \includegraphics[width=0.9\columnwidth]{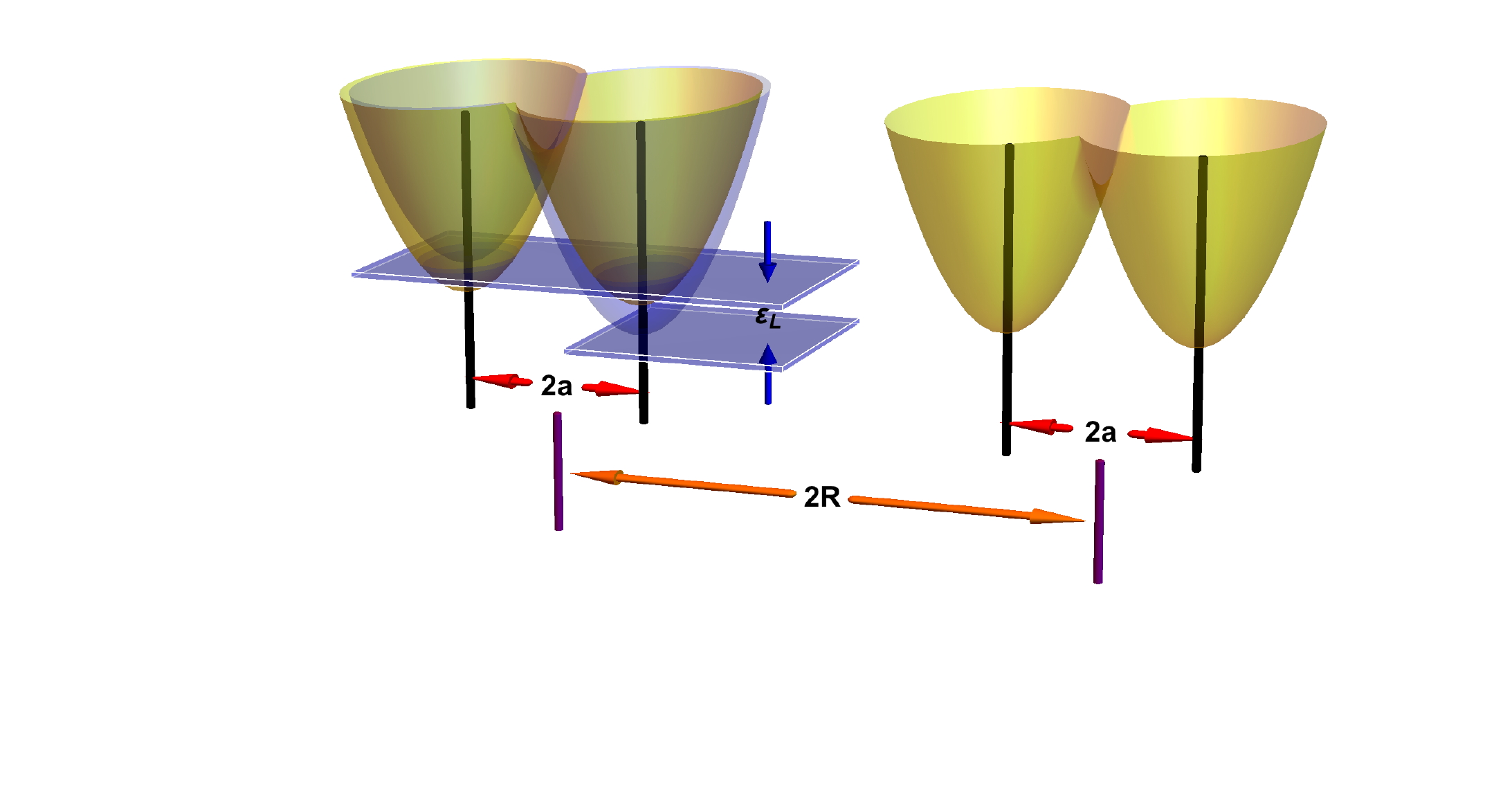}
\caption{Schematic diagram of a lateral four-quantum-dot system hosting two singlet-triplet qubits. The parameters $2a$ (the distance between the center of two adjacent dots within one double dot system),  $2R$ (the distance between the center of two double-dot systems), and  $\varepsilon_L$ (the detuning on the left double dots) are indicated.}
\label{fig:1}
\end{figure}

\section{Model and Methods}
\label{sec:model}
We consider a lateral four-quantum-dot system under a homogeneous magnetic field $\bm{B}$  along the $z$-axis. The Hamiltonian  can be generally written, in the $xy$ plane, as
\begin{equation}
H=\sum_i\left[\frac{1}{2m^*}\left(\bm{p}_i-e\bm{A}\right)^2+V(\bm{r}_i)\right],
\label{eq:H}
\end{equation}
where $m^*$ is the effective mass of the electron. $V(\bm{r})$ is the confinement potential of four quantum dots, consisted of a pair of Double Quantum Dots (DQDs) separated by a distance $2R$, while the interdot distance within each DQD is $2a$. In our work we model $V(\bm{r})$ with the form 
\begin{equation}
\begin{aligned}
V(\bm{r})=\text{Min}\left[v_1(\bm{r}),v_2(\bm{r}),v_3(\bm{r}),v_4(\bm{r})\right],
\end{aligned}
\label{eq:V}
\end{equation}
where
\begin{equation}
\begin{aligned}
v_i(\bm{r})\equiv\frac{m^*\omega_0^2}{2}\left|\bm{r}-\bm{R}_i\right|^2+\epsilon_i
\end{aligned}
\label{eq:v}
\end{equation}
indicates the confinement potential of the $i$th quantum dot centering at $\bm{R}_i$, the coordinates of which are 
\begin{equation}
\begin{aligned}
\bm{R}_1=(-R-a,\text{ }0),&\quad\bm{R}_2=(-R+a,\text{ }0),\\
\bm{R}_3=(R-a,\text{ }0),&\quad\bm{R}_4=(R+a,\text{ }0).
\end{aligned}
\label{eq:R}
\end{equation}
In Eq.~\eqref{eq:v}, $\omega_0$ is the confinement energy\cite{Burkard.99, Hu.00} that characterizes the size of the dot; differences in $\epsilon_i$'s are used to represent the detuning of the corresponding DQD.
A schematic diagram of the potential is shown in Fig.~\ref{fig:1}. In most cases we have considered in this work, there is only one electron in each dot. In this case, each DQD can be treated as one singlet-triplet (S-T) qubit, the detuning of which is represented by $\varepsilon_L$ and $\varepsilon_R$ for the left S-T qubit (dots 1 and 2) and the right one (dots 3 and 4) respectively,\cite{Petta.05}
i.e. $\varepsilon_L=\epsilon_1-\epsilon_2$ and $\varepsilon_R=\epsilon_3-\epsilon_4$.

We adopt the Hund-Mulliken method to solve this multi-electron problem.\cite{Burkard.99,Hu.00} The starting point is the Fock-Darwin states, which are the electron wave functions in an isolated dot modeled by a harmonic potential. Among them only the ground states
\begin{equation}
\begin{aligned}
\phi_i(\bm{r})=\frac{1}{a_B^{}\sqrt{\pi}}\text{exp}\left[{-\frac{1}{2a_B^2}\left|\bm{r}-\bm{R}_i\right|^2}\right],\quad i=1,2,3,4,
\end{aligned}
\label{eq:ground}
\end{equation}
are retained in the Hund-Mulliken approximation.
Here, $a_B^{}\equiv\sqrt{\hbar/m\omega_0}$ is Fock-Darwin radius. We then build a set of orthogonal single-electron states by the transformation
\begin{equation}
\begin{aligned}
\left\{\psi_1,\text{ }\psi_2,\text{ }\psi_3,\text{ }\psi_4\right\}^\text{T}=\mathcal{O}^{-1/2}\left\{\phi_1,\text{ }\phi_2,\text{ }\phi_3,\text{ }\phi_4\right\}^\text{T},
\end{aligned}
\label{eq:trans}
\end{equation}
where $\mathcal{O}$ is the overlap matrix defined as $\mathcal{O}_{l,l^\prime}\equiv\langle\phi_l|\phi_{l^\prime}\rangle$. There are different ways to obtain  $\mathcal{O}^{-1/2}$ in the literature, e.g. in Ref.~\onlinecite{Barnes.11} a Gram-Schmidt method has been performed for the orthogonalization. However, In our situation we would like to keep the symmetry that the states remain the same after indices 1,4 and 2,3 are interchanged ($1\leftrightarrow4$ and $2\leftrightarrow3$). We therefore used the method by L\"owdin's orthogonalization 
where the $\mathcal{O}^{-1/2}$ can be obtained by the singular value decomposition.\cite{annavarapu2013singular} We note that Eq.~\eqref{eq:trans} marks the main difference of this work from previous ones.\cite{Calderon.15} In previous works dealing with four quantum dots, the two DQDs are treated essentially separately and $\psi_{1,2}$ are merely dependent on $\phi_1$ and $\phi_2$ while  $\psi_{3,4}$ on $\phi_3$ and $\phi_4$. Here, as an example, our $\psi_1$ is explicitly an admixture of all $\phi_1$ through $\phi_4$. The consequences of this difference shall be clear later.

In this work we consider four electrons (two spin up and two spin down) occupying the lateral four-quantum-dot system. Since only the lowest energy levels for each dot are retained in the Hund-Mulliken approximation, the maximum occupancy for each dot is two. The four-electron wave function can be generically written as 
\begin{equation}
\begin{aligned}
|\Psi_j\rangle=c^\dagger_{m\uparrow}c^\dagger_{n\downarrow}c^\dagger_{p\uparrow}c^\dagger_{q\downarrow}|\text{vac}\rangle,\end{aligned}
\label{eq:basis}
\end{equation}
where $c^\dagger_{k,\sigma}$ creates an electron with spin $\sigma$ at the $k^{\rm th}$ dot, $m,n,p,q\in\{1,2,3,4\}$,  $j$ labels different energy levels for the four-electron states, and $|\text{vac}\rangle$ describes a vacuum state. The four-electron Hamiltonian of the problem can then be expressed in a Hubbard-like form as is shown in the Appendix as Eq.~\eqref{eq:fqdHam}. There are 36 possibilities for the four electrons to occupy the lowest levels of four dots (two spin up electrons see four possibilities and so are the two spin down ones), therefore the Hamiltonian under the bases Eq.~\eqref{eq:basis} is a $36\times36$ matrix. In the Hubbard form of the Hamiltonian \eqref{eq:fqdHam}, the kinetic energy ($H_e$) and Coulomb repulsion ($H_U$) terms constitute the diagonal terms of the matrix and the hopping ($H_t$), spin super-exchange ($H_{J^e}$), pair-hopping ($H_{J^p}$) and occupation-modulated hopping ($H_{J^t}$) terms primarily contribute to the off-diagonal ones.
In practice, the matrix elements of the Hamiltonian are evaluated by computing the overlaps of the corresponding wave functions, that is effectively replacing the creation operators in Eq.~\eqref{eq:basis} by the orthogonalized single-particle wave functions $\psi_1$ though $\psi_4$ defined in Eqs.~\eqref{eq:trans}. 
In other words, the parameters of the Hamiltonian \eqref{eq:fqdHam} are calculated microscopically using the Hund-Mulliken wave functions \eqref{eq:trans}. One can then use either the matrix form or the second quantized form of the Hamiltonian to further study its physical properties, including the energy spectrum and the (effective) exchange interaction as we will discuss in the remainder of this paper.

In a previous work,\cite{Calderon.15} this problem was treated in the case where the two pairs of double-quantum-dot systems are sufficiently far away from each other such that the Hund-Mulliken wave function for each S-T qubit does not involve components from the other one. In other words, in Ref.~\onlinecite{Calderon.15} only the Coulomb repulsion is considered between the second and third electrons, and hopping and exchange involving them are ignored. This has led to a dramatic simplification of the problem as the Hilbert space can be block-diagonalized and the two S-T qubits are treated separately. In contrast, our work explicitly retains all interactions involving the second and third electrons and our Hund-Mulliken wave functions include contributions from all four electrons. 

\section{Results}
\label{sec:res}

\subsection{Energy spectrum}
\label{sec:spectrum}

In this section, we study the energy spectrum of the four-quantum-dot system described in the previous section. To begin with, we qualitatively describe the situation where the two S-T qubits are sufficiently far away such that the exchange interaction between them can be ignored, as that provides the first insight to the problem at hand. In this case, the Hamiltonian can be factorized into two parts describing the left (``L'') and right (``R'') S-T qubits respectively: $H=H^\text{L}\otimes H^\text{R}$, while the states relevant to quantum computation can be written as $|SS\rangle$, $|ST\rangle$, $|TS\rangle$, and $|TT\rangle$, where the first entry in the ket indicates the state in the left S-T qubit and the second the right one. To facilitate the discussion we introduce the following notations describing a state of an S-T qubit: (we have taken the left S-T qubit involving dots 1 and 2 as the example, but one can easily replace them by dots 3 and 4 as well.)
\begin{subequations}
\begin{align}
|S(11)\rangle&=\frac{1}{\sqrt{2}}\left(c^\dagger_{1\uparrow}c^\dagger_{2\downarrow}-c^\dagger_{1\downarrow}c^\dagger_{2\uparrow}\right)|\text{vac}\rangle,\\
|S(20)\rangle&=c^\dagger_{1\uparrow}c^\dagger_{1\downarrow}|\text{vac}\rangle,\\
|S(02)\rangle&=c^\dagger_{2\uparrow}c^\dagger_{2\downarrow}|\text{vac}\rangle,\\
|T(11)\rangle&=\frac{1}{\sqrt{2}}\left(c^\dagger_{1\uparrow}c^\dagger_{2\downarrow}+c^\dagger_{1\downarrow}c^\dagger_{2\uparrow}\right)|\text{vac}\rangle.
\end{align}
\end{subequations}
For later use we also define a four electron state as
\begin{align}
|S(\uparrow\uparrow\downarrow\downarrow)\rangle&=\frac{1}{\sqrt{2}}\left(c^\dagger_{1\uparrow}c^\dagger_{2\uparrow}c^\dagger_{3\downarrow}c^\dagger_{4\downarrow}-c^\dagger_{1\downarrow}c^\dagger_{2\downarrow}c^\dagger_{3\uparrow}c^\dagger_{4\uparrow}\right)|\text{vac}\rangle.
\end{align}

\begin{figure}
  \includegraphics[width=1\columnwidth]{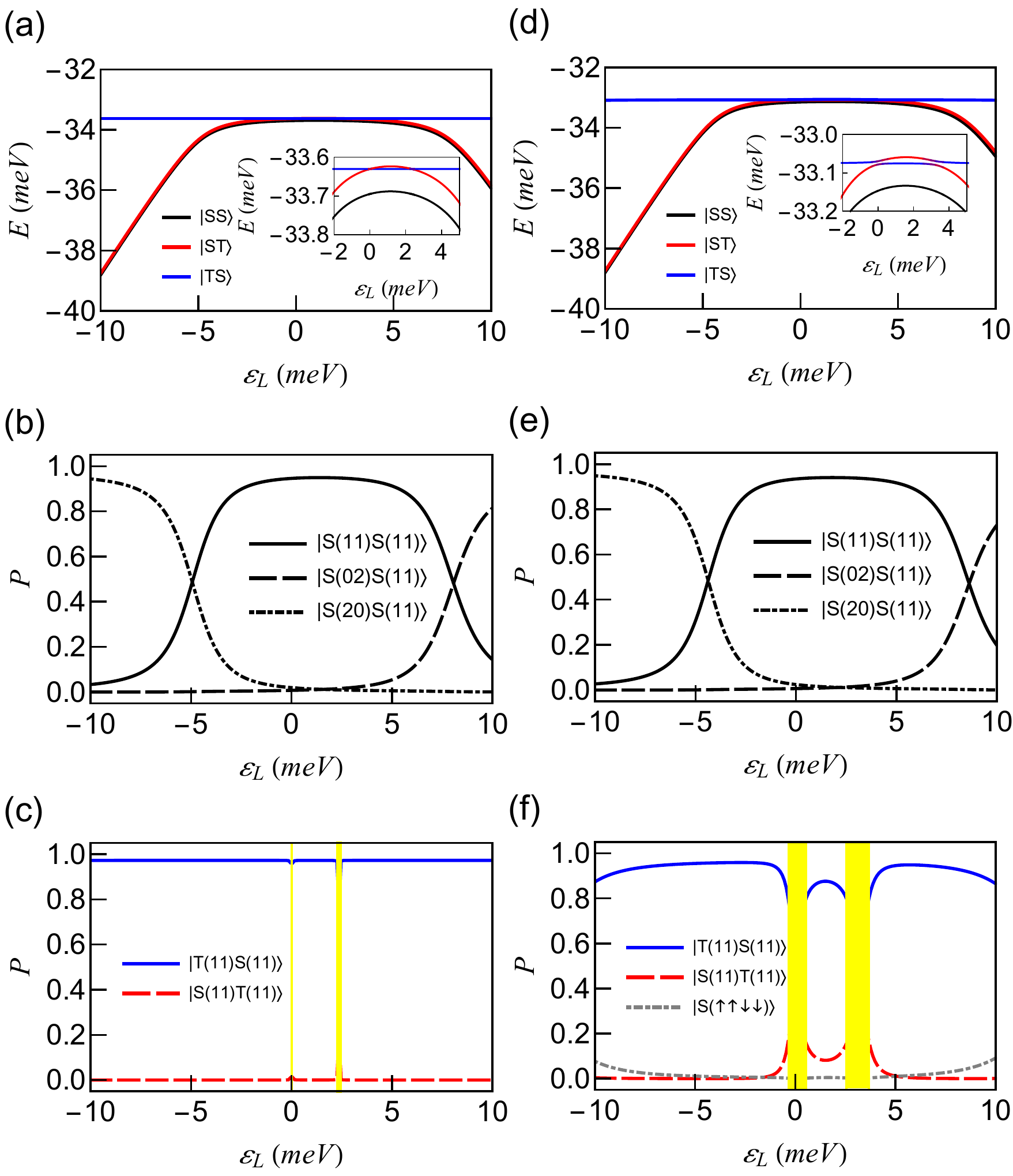}
  \caption{Energy spectra and the state composition as functions of the detuning $\varepsilon_L$. (a)-(c) (left column): $R=58 \text{nm}$. (d)-(f) (right column): $R=49 \text{nm}$. Common parameters: $\hbar\omega_0=5 \text{meV}$, $a=22 \text{nm}$.  (a) and (b) show the three energy levels relevant to the manipulation of the left S-T qubit. The blue(horizontal) lines show the energy for the  $|TS\rangle$ state, while the black and red(gray) lines show the energy for the  $|SS\rangle$ and $|ST\rangle$ states respectively. The latter two are very close and can only be distinguished on a finer scale (shown in the insets).  (c) and (d) show the composition of the $|SS\rangle$ state, while (e) and (f) show the composition of the $|TS\rangle$ state, with the $y$-axis label $P$ as the probability of each state indicated on the figure.}\label{fig:2}
\end{figure}

Fig.~\ref{fig:2}(a) shows the energy spectrum of three levels relevant to the manipulation of the S-T qubits out of the full spectrum, with $R=2.6a$. Fig.~\ref{fig:2}(b) shows the components of the $|SS\rangle$ state and Fig.~\ref{fig:2}(c) those of the $|TS\rangle$ state as indicated in Fig.~\ref{fig:2}(a).
In this case, the two pairs of double quantum dots are sufficiently far away that the result closely resembles the case in which they behave independently.  We have also verified that for all $R>2.6a$ the results are very similar to the case with two independent S-T qubits, but those results are not shown. We see that if both S-T qubits have zero detuning, i.e. all dots are leveled, the lowest energy state is the $|SS\rangle$ state, but it is very close to the other states, $|ST\rangle$ and $|TS\rangle$ as shown in the figure. (The $|TT\rangle$ state is not important for the manipulation of the left qubit thus is not shown here.)  The $|SS\rangle$ state has mostly the $|S(11)S(11)\rangle$ character while the $|TS\rangle$ state almost 100\% $|T(11)S(11)\rangle$.
Detuning of the left qubit (changing $\varepsilon_L$) leads to a reduction in energy of the $|SS\rangle$ state and a increase of the portion of $|S(20)S(11)\rangle$ ($\varepsilon_L<0$) or $|S(02)S(11)\rangle$ ($\varepsilon_L>0$) in its composition, while the $|TS\rangle$ state remain unchanged. The energy difference between the $|SS\rangle$ and $|TS\rangle$ states therefore gives the exchange interaction required to steer the left qubit, performing a rotation of the Bloch vector around the $z$ axis of the Bloch sphere. These are known results from previous literature.\cite{Burkard.99, Hu.00, Li.10, Calderon.15}

The inset of Fig.~\ref{fig:2}(a) presents a zoomed-in view of the three levels near the zero detuning. It is interesting to note that the two levels $|ST\rangle$ and $|TS\rangle$ cross at $\varepsilon_L\approx 0$ and $\varepsilon_L\approx 2.4$ meV. The two crossings are actually very sharp avoided crossings, which are direct consequences of the interactions between the second and third electrons belonging to the left and right S-T qubit respectively. In particular, the spin super-exchange term results in a nonzero off-diagonal elements between the  $|ST\rangle$ and $|TS\rangle$ states, and such avoided crossing is just a manifestation that the two S-T qubits are not infinitely far away. Such avoided crossings are very sharp and only affect a small range of $|TS\rangle$ as a function of $\varepsilon_L$. These are noted as yellow  lines in Fig.~\ref{fig:2}(c).

As the two S-T qubits become closer, the mixture of $|ST\rangle$ and $|TS\rangle$ states at certain detuning values becomes more pronounced. Fig.~\ref{fig:2}(d)-(f) show a situation with $R/a\approx2.2$. The first difference we notice between Fig.~\ref{fig:2}(a) and (d) is that the curves are moved in entirety to the right in (d) compared to (a). This is because as the two S-T qubits become closer, the Coulomb repulsion imposed by the right qubit to the left qubit becomes stronger, which increases the tendency to have double occupancy in the leftmost dot (the first dot). In other words, some of the Coulomb repulsion takes up the role of the detuning and now a weaker detuning is needed to achieve the same level of exchange interaction compared to the case in which the two S-T qubits are far away. However, the most important feature we notice from the inset of Fig.~\ref{fig:2}(d) is that the avoided crossings between the $|ST\rangle$ and $|TS\rangle$ states are now much more visible and are smoother, and the percentage of the expected qubit states in the actual admixture of $|SS\rangle$ or $|TS\rangle$ states are reduced. Fig.~\ref{fig:2}(e) shows the composition of the $|SS\rangle$ state as a function of the detuning and the percentage of $|S(1,1)S(1,1)\rangle$ state for the small detuning regime are not decreased much. On the contrary, the composition of the $|TS\rangle$ state has been appreciably changed. From Fig.~\ref{fig:2}(f) we notice that a new state, the $|S(\uparrow\uparrow\downarrow\downarrow)\rangle$ state now enters the admixture as a result of the spin super-exchange between the second and third spins, and most importantly in the neighborhood of the avoided crossing $|T(11)S(11)\rangle$ and $|S(11)T(11)\rangle$ states are substantially mixed. Since $|T(11)S(11)\rangle$ no longer dominates the state, one is unable to define an exchange interaction in the usual way of taking the energy difference between the state and the $|SS\rangle$, leading to difficulties in understanding the manipulation of multi-quantum-dot systems. In Fig.~\ref{fig:2}(f) we have masked the range where the portion of $|T(11)S(11)\rangle$ drops below 80\% with two yellow (gray) rectangles, and one must exercise caution in defining the exchange interaction as the energy level splitting between the two states concerned. 

Overall the results of the energy spectra of the four dot systems show interesting features: When the two pairs are sufficiently far away, the results closely resemble that of two individual S-T qubits. On the other hand as they get closer the spin super-exchange between the second and third electrons introduces the $|S(\uparrow\uparrow\downarrow\downarrow)\rangle$ state into the composition of the $|TS\rangle$ state, and mixes
the $|T(11)S(11)\rangle$ and $|S(11)T(11)\rangle$ states near two points of the avoided crossing of energy spectra. These results motivate us to study how an effective exchange interaction may be defined in the next section.

\begin{figure}
  \includegraphics[width=1\columnwidth]{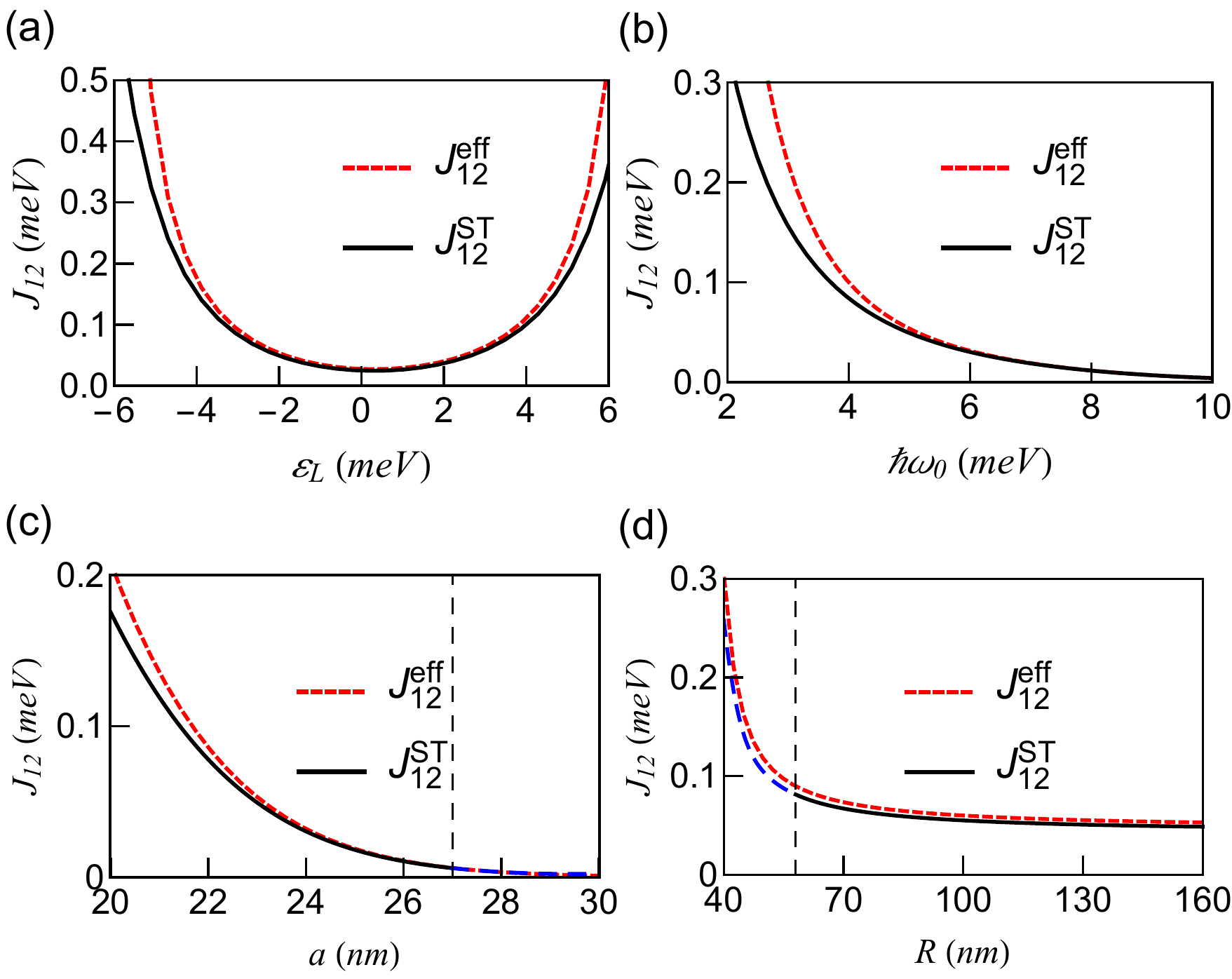}
  \caption{Comparison of the exchange interaction calculated from the energy difference between the $|SS\rangle$ and $|TS\rangle$ states, $J_{12}^\mathrm{ST}$ (the black solid lines), and the effective  exchange interaction $J_{12}^\mathrm{eff}$ derived from Eq.~\eqref{eq:spinchain1} (the red(gray) dashed lines). (a) $J_{12}$ as functions of detuning $\varepsilon_L$. $\hbar\omega_0=5 \text{meV}$, $a=22 \text{nm}$, and 
  $R=150 \text{nm}$.
  (b) $J_{12}$ as functions of the confinement energy $\hbar\omega_0$. $a=22 \text{nm}$, $R=150 \text{nm}$, and $\varepsilon_L=-2 \text{meV}$.
  (c) $J_{12}$ as functions of $a$. $\hbar\omega_0=5 \text{meV}$, $R=60 \text{nm}$, and $\varepsilon_L=-2 \text{meV}$.
  (d) $J_{12}$ as functions of $R$.  $\hbar\omega_0=5 \text{meV}$, $a=22 \text{nm}$, and $\varepsilon_L=-2 \text{meV}$. The blue(gray) long dashed line between $R=40$nm and $R=58$nm (shown as a vertical black dashed line) indicates a range where $J_{12}^\mathrm{ST}$ is ill defined.}\label{fig:3}
\end{figure}

\subsection{Exchange interaction}
\label{sec:exch}

In this section we present results of the effective exchange interaction of the S-T qubit with another S-T qubit nearby, and compare the results to those calculated by taking the energy difference between the two computational states, where applicable. The lateral four-quantum-dot system can be alternatively described by a Heisenberg spin chain model, \cite{nagaosa1999quantum}
\begin{equation}
\begin{aligned}
H^\mathrm{eff}=J_{12}^\mathrm{eff}\bm{S}_1\cdot\bm{S}_2+J_{23}^\mathrm{eff}\bm{S}_2\cdot\bm{S}_3+J_{34}^\mathrm{eff}\bm{S}_3\cdot\bm{S}_4.
\end{aligned}
\label{eq:spinchain}
\end{equation}
where $\bm{S}_i$ is the spin operator for the electron occupying the $i$th dot, and $J_{ij}^\mathrm{eff}$ the effective exchange interaction between electrons in the $i$th and $j$th dot. The spin Hamiltonian \eqref{eq:spinchain} is related to the Hubbard-like Hamiltonian \eqref{eq:fqdHam} by a Schrieffer-Wolff transformation.\cite{Schrieffer.66,Bravyi.11, Li.12} Detailed derivations are given in the Appendix, where we also show that the transformation gives the effective exchange interactions as functions of the Hubbard parameters as
\begin{equation}
\begin{aligned}
J^{\text{eff}}_{12}=&\frac{2(t_{12}-J^{t1}_{12})^2}{U_1-U_{12}-U_{23}+\epsilon_1-\epsilon_2}\\
&+\frac{2(t_{12}-J^{t2}_{12})^2}{U_2-U_{12}+U_{23}-\epsilon_1+\epsilon_2}-2J^e_{12},
\end{aligned}
\label{eq:spinchain1}
\end{equation}
\begin{equation}
\begin{aligned}
J^{\text{eff}}_{23}=&\frac{2(t_{23}-J^{t2}_{23})^2}{U_2+U_{12}-U_{23}-U_{34}+\epsilon_2-\epsilon_3}\\
&+\frac{2(t_{23}-J^{t3}_{23})^2}{U_3+U_{34}-U_{12}-U_{23}-\epsilon_2+\epsilon_3}-2J^e_{23},\\
\end{aligned}
\label{eq:spinchain2}
\end{equation}
and
\begin{equation}
\begin{aligned}
J^{\text{eff}}_{34}=&\frac{2(t_{34}-J^{t3}_{34})^2}{U_3+U_{23}-U_{34}+\epsilon_3-\epsilon_4}\\
&+\frac{2(t_{34}-J^{t4}_{34})^2}{U_4-U_{23}-U_{34}-\epsilon_3+\epsilon_4}-2J^e_{34}.\\
\end{aligned}
\label{eq:spinchain3}
\end{equation}

Since the Hubbard parameters can always be calculated from the overlap integrals of the wave functions, we can calculate the effective exchange interactions for the cases even if the two S-T qubits are close to each other and the desired computational states are substantially admixed with others. To benchmark whether this is a judicious definition of the exchange interaction, we compare the effective exchange interaction $J^{\text{eff}}_{12}$ to that calculated from the energy level differences, $J^{\text{ST}}_{12}$. The results are shown in Fig.~\ref{fig:3}. Figure~\ref{fig:3}(a) shows $J^{\text{eff}}_{12}$ and  $J^{\text{ST}}_{12}$ as functions of the detuning $\varepsilon_L$ with fixed confinement energy and the location of the dots. The results for $J^{\text{eff}}_{12}$ and  $J^{\text{ST}}_{12}$ agree reasonably well for small detuning ($|\varepsilon_L|\lesssim2\text{meV} $), and they both show the correct trend: the exchange interaction is minimal when $\varepsilon_L=0$, and it increases when the left S-T qubit is detuned in both directions, either $\varepsilon_L<0$ or $\varepsilon_L>0$. For larger absolute value of $\varepsilon_L$, $J^{\text{eff}}_{12}$ tends to overestimate the exchange interaction, which is about 20\% larger than $J^{\text{ST}}_{12}$ at $|\varepsilon_L|=5\text{meV}$. Nevertheless, taking into account of the fact that the manipulation of an S-T qubit is done at small detuning values, i.e. the singlet state should not be dominated by $|S(2,0)\rangle$ or $|S(0,2)\rangle$ states, we conclude that both methods of calculating the exchange interaction agree within this range.

Figure~\ref{fig:3}(b) shows $J^{\text{eff}}_{12}$ and  $J^{\text{ST}}_{12}$ as functions of the confinement energy $\hbar\omega_0$ with the detuning $\varepsilon_L=-2\text{meV}$ and the location of the dots fixed. Both lines show clear trend that for larger $\hbar\omega_0$ (corresponding to narrower dots and higher potential barriers in between) the exchange interactions are smaller, while for smaller $\hbar\omega_0$ (corresponding to wider dots and lower potential barriers)  the exchange interactions are larger.  $J^{\text{eff}}_{12}$ and $J^{\text{ST}}_{12}$ basically agree for dots that are sufficiently narrow, i.e. $\hbar\omega_0\ge5\text{meV}$, but for wider dots $J^{\text{eff}}_{12}$ becomes systematically larger than $J^{\text{ST}}_{12}$. We note that when the dots are wide and the potential barriers in between are low, the wave functions between the two dots are substantially mixed and the Hund-Mulliken approximation becomes less reliable. Nevertheless, both curves show the same trend and the deviation between the two are not large, provided the fact that in realistic experiment the potential barriers between the dots are typically well defined.

Figure~\ref{fig:3}(c) shows $J^{\text{eff}}_{12}$ and  $J^{\text{ST}}_{12}$ as functions of the half distance $a$ between the two dots within each S-T qubit, with the confinement energy $\hbar\omega_0=5\text{meV}$ and the distance between the two pairs of DQD $2R=120$ nm fixed. For small $a$, the two dots within each S-T qubit are close to each other and the exchange interaction, which is essentially the overlap of the electron wave functions in the two dots, are large. As $a$ is increased, the exchange interaction rapidly decreases. This trend has been clearly shown in Fig.~\ref{fig:3}(c). We again encounter a similar situation with the other panels that $J^{\text{eff}}_{12}$ tends to overestimate compared to $J^{\text{ST}}_{12}$. However, an important message from this figure is that $J^{\text{eff}}_{12}$ can be defined in a wider domain than $J^{\text{ST}}_{12}$. In particular, when $a\ge27\text{nm}$, the second and the third dots, which belongs to two S-T qubits respectively, becomes close enough to each other that the computational states no longer dominate the four electron wave functions, and the exchange interaction therefore cannot be extracted as the energy difference between the two relevant states. This fact is signified in Fig.~\ref{fig:3}(c) by a dashed vertical line at $a=27\text{nm}$, beyond which $J^{\text{ST}}_{12}$ is shown as the long dashed line. Nevertheless, we may still define the effective exchange interaction according to Eq.~\eqref{eq:spinchain1}, and it turns out that it agrees well with the results of $J^{\text{ST}}_{12}$ that would be obtained if one simply takes the energy difference. This fact means that the  $J^{\text{ST}}_{12}$ obtained from taking the energy difference between the two states relevant to quantum computation, even when they are admixtures of a set of states without being dominated by the desired computational states, can be regarded as the effective exchange interaction.

Figure~\ref{fig:3}(d) compares the variation of $J^{\text{eff}}_{12}$ and  $J^{\text{ST}}_{12}$ as the distance between the S-T qubits $2R$ is changed, with the confinement energy $\hbar\omega_0=5\text{meV}$, the detuning $\varepsilon_L=-2\text{meV}$ and $a=22\text{nm}$ fixed. For $R$ sufficiently large, the two S-T qubits behave essentially independently, and both curves of the exchange interaction show little changes for $R\gtrsim100\text{nm}$ as expected. As $R$ is reduced, the two S-T qubits become closer and that leads to an increase in the exchange interaction. The reason is as explained in Fig.~\ref{fig:2}: the right S-T qubit imposes an additional Coulomb repulsion to the left S-T qubit, and that has increased the portion of $|S(2,0)\rangle$ in the singlet state, which has effectively enhanced the exchange interaction. Both curves show this trend as expected. For $R<58\text{nm}$, the two S-T qubits are so close to each other that again the computational states no longer dominate the actual four electron state. A vertical dashed line is drawn at this point and $J^{\text{ST}}_{12}$ is marked as dashed line to the left of it. Nevertheless,  $J^{\text{eff}}_{12}$, albeit still overestimate a little, shows that one may still define effective exchange interactions in this situation.

\subsection{Entangling operations and the gate crosstalk}\label{sec:crosstalk}

Based on the above results, we now move on to study the operations of two S-T qubits. There are two ways to perform entangling operations between a pair of S-T qubits. The first method makes use of the exchange interaction between the neighboring spins belonging to different qubits (i.e. $J^{\text{eff}}_{23}$ in this paper), and the second method uses the capacitive coupling created when the qubits are detuned such that some of the quantum dots become doubly occupied. The latter mechanism can be summarized by the following Hamiltonian:\cite{Ramon.11,Calderon.15}
\begin{equation}
\begin{split}
H_{\rm int}=&-\beta_1(\varepsilon_L,\varepsilon_R)\sigma_z\otimes I
-\beta_2(\varepsilon_L,\varepsilon_R)I\otimes\sigma_z\\
&+\alpha_0(\varepsilon_L,\varepsilon_R)\sigma_z\otimes\sigma_z,\label{eq:Hcapac}
\end{split}
\end{equation}
where $\beta_{1,2}$ and $\alpha_0$ can be found by Coulomb integrals as prescribed in Refs.~\onlinecite{Ramon.11} and \onlinecite{Calderon.15}.

It is known that the exchange coupling $J^{\text{eff}}_{23}$ between the two qubits, relying on the overlap between wave functions, decays exponentially as the qubits become far apart. Therefore, in this case the capacitive coupling shall dominate. On the other hand, as the two qubits are drawn closer to each other, while both the exchange and capacitive coupling increase, the exchange coupling is expected to increase more rapidly and will eventually be stronger than the capacitive coupling when the two qubits are close enough.

\begin{figure}
  \includegraphics[width=0.7\columnwidth]{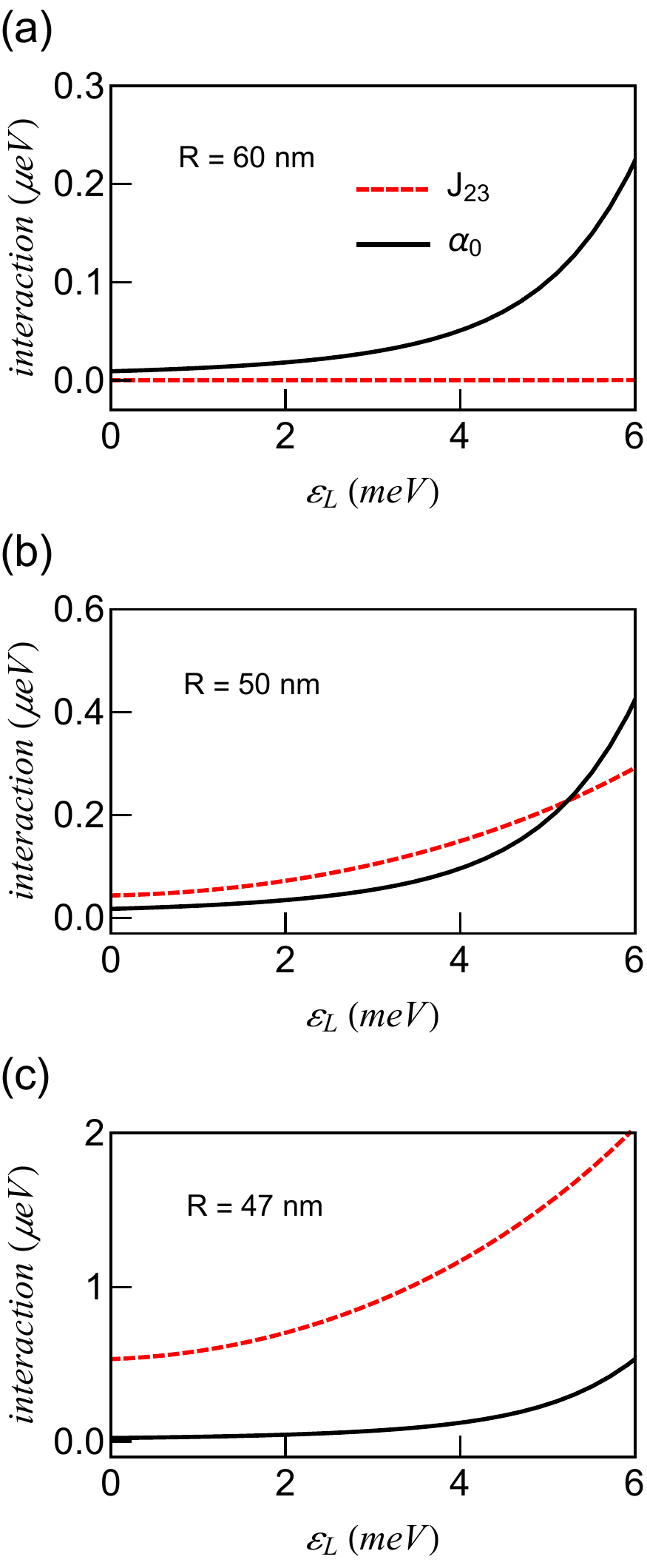}
  \caption{Comparison of calculated $J_{23}^{\rm eff}$ (black lines) and $\alpha_0$ (red(gray) dashed lines) as functions of the detuning of the left qubit $\varepsilon_L$, while $\varepsilon_R$ is maintained to be zero. (a) $R=60$ nm. (b) $R=50$ nm. (c) $R=47$ nm. Common parameters: $\hbar\omega_0=5 \text{meV}$, $a=22 \text{nm}$.}\label{fig:4}
\end{figure}

The results presented in the previous sections of this paper allow us to quantitatively compare the exchange and capacitive coupling for two S-T qubits at certain distance. We focus on the parameter of the entangling term $\alpha_0$. The comparison of calculated $J_{23}^{\rm eff}$ and $\alpha_0$ values as functions of the detuning of the left qubit $\varepsilon_L$ is shown in Fig.~\ref{fig:4}. $\varepsilon_R$ is maintained to be zero. From Fig.~\ref{fig:4}(a), we see that the exchange interaction $J_{23}^{\rm eff}$ is essentially zero for $R/a\approx 2.73$ because the two S-T qubits are too far away, as expected. On the other hand, the capacitive coupling $\alpha_0$ clearly increases with the detuning of the left qubit. As the two S-T qubits move closer, both the exchange and capacitive interaction increase in magnitude, but the increase of the exchange interaction is more pronounced. For $R/a\approx 2.27$ [Fig.~\ref{fig:4}(b)] the two are already roughly comparable, and as they come even closer [cf. Fig.~\ref{fig:4}(c) for results at $R/a\approx 2.17$] the exchange interaction increase so rapidly that it greatly exceeds the capacitive interaction.

\begin{figure}
  \includegraphics[width=0.8\columnwidth]{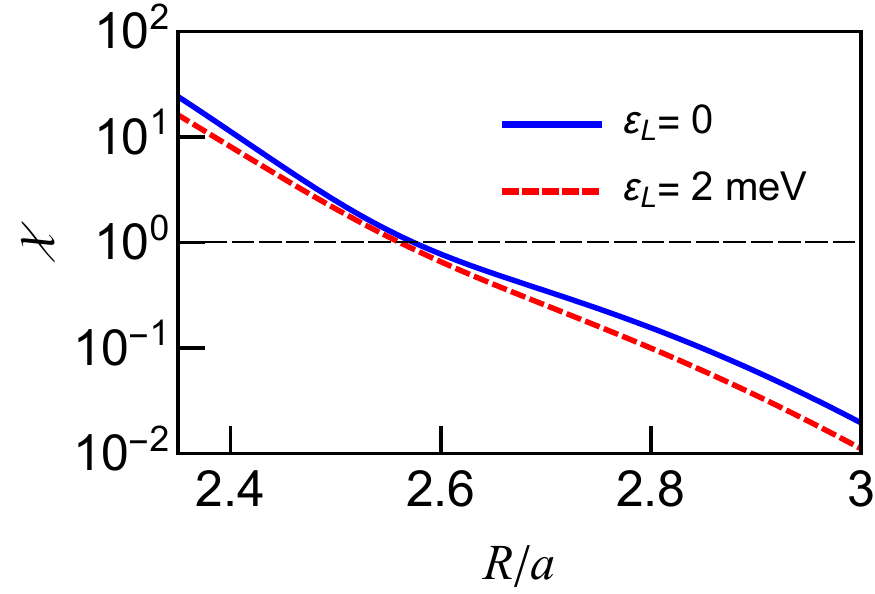}
  \caption{Ratio between the exchange and capacitive interactions,  $\chi=J_{23}^{\rm eff}/\alpha_0$, as functions of $R/a$ for two different detuning values as indicated.Parameters: $\hbar\omega_0=5 \text{meV}$, $a=22 \text{nm}$.}\label{fig:5}
\end{figure}

The results presented in Fig.~\ref{fig:4}(b) suggest that while the evolution of a pair of coupled S-T qubit system can be regarded as dominated by the capacitive interaction when the two qubits are far apart, or by the exchange interaction when they are close-by, it is actually more subtle that for two qubits at intermediate distances the two coupling schemes are comparable in strength, complicating the analyses of the evolution of the two-qubit system. We have taken the ratio between  $J_{23}^{\rm eff}$ and $\alpha_0$ (denoted as $\chi$) at two different detuning values $\varepsilon_L=0$ and 2 meV and the results are shown in Fig.~\ref{fig:5}. Both lines in Fig.~\ref{fig:5} cross $\chi=1$ at approximately the same place, $R/a\approx 2.58$, and this distance is relevant to experiments. The results shown in Fig.~\ref{fig:5} suggests that in order to determine whether a pair of exchange-coupled S-T qubits be coupled predominately by the exchange or capacitive interaction, attention must be paid to the distance between them so that it can be fabricated either considerably less than or greater than the ``intermediate'' distance $R/a\approx 2.58$, so that either the exchange or capacitive coupling, but not both, shall dominate the evolution.

Another interesting aspect arising in the operation of coupled four-quantum-dot system is the gate crosstalk, a situation that when an individual qubit is being addressed, its adjacent qubits are affected even if they have been left alone. The capacitive coupling between the two S-T qubits essentially forms a kind of such gate crosstalk because as one of the qubits are detuned, the singlet-triplet energy splitting of the other is changed. This effect is encapsulated in the parameters $\beta_{1,2}$ of Eq.~\eqref{eq:Hcapac}.

The gate crosstalk may also manifest itself in two other channels.
In the first channel, the changes in gate voltages are distributed into different dots. Of course, if one is addressing one particular dot with certain gate voltage, that voltage will have a major effect on the chemical potential of the electrons therein. Nevertheless, the energy of electrons in adjacent dots may also be shifted and the linear combination of these energy changes together accounts for the gate voltage applied. Since this is entirely electrostatic, we term this channel as the ``classical'' one. The phenomenon has been studied since the early development of quantum dot spin qubits and is well explained by the capacitance circuit model,\cite{Hanson.07} which essentially keeps all the kinetic energy, Zeeman and Coulomb repulsion terms in the Hamiltonian \eqref{eq:fqdHam}, namely terms in Eqs.~\eqref{eq:A6a} and \eqref{eq:A6c} are retained.\cite{Yang.11} This effect can be straightforwardly compensated, for example, by using appropriate linear combinations of relevant gate voltages and we therefore will not cover this further in this work.

We will therefore focus on the ``quantum'' channel of the gate crosstalk, which stems completely from the overlap between the wave functions of the adjacent electrons belonging to different S-T qubits. This effect is expectedly much weaker than the classical channel but is more difficult to compensate. Therefore one of the main purposes of this paper is to give a quantitative evaluation of the deteriorating effect of the quantum aspect of the gate crosstalk, which we refer to as simply the ``gate-crosstalk''. We note that in previous works treating the four dot system as two separate pairs, the gate crosstalk is absent because there will be no change on the quantum state of a given S-T qubit when the other one is being manipulated. In our work, we take into account of all four electron states and we can capture the gate crosstalk accordingly. The gate crosstalk manifests itself as the change in the exchange interaction of an idle S-T qubit while another is being addressed. We expect that this effect will be minimal when the two S-T qubits are far away from each other, but will become more pronounced when they are close. As mentioned in Sec.~\ref{sec:exch}, the usual way of defining the exchange interaction as the energy difference between the qubit levels becomes problematic when $R/a\lesssim2$. Nevertheless, the effective exchange interaction $J^\text{eff}$ defined in Eqs.~\eqref{eq:spinchain1} and \eqref{eq:spinchain3} has always been well-defined, and while $J^\text{eff}$ tends to overestimate, they agree well in the parameter range where quantum computation is to be performed. We will therefore use $J^\text{eff}$ to study the gate crosstalk.

\begin{figure}
  \includegraphics[width=1\columnwidth]{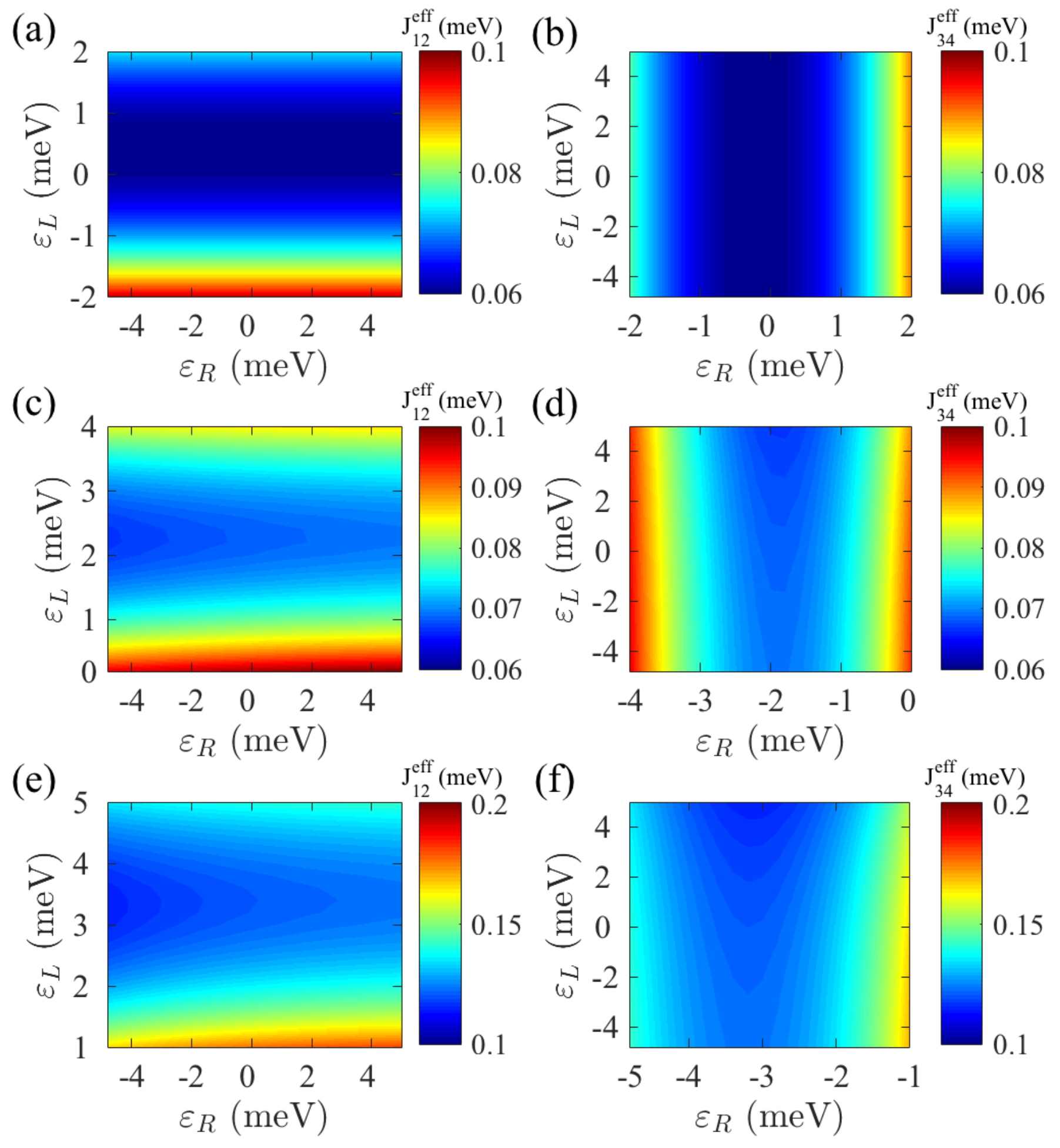}
  \caption{Pseudo-color plot of the effective  exchange interaction $J_{12}^\mathrm{eff}$ [(a), (c) and (e)] and $J_{34}^\mathrm{eff}$ [(b), (d) and (f)] derived from Eqs.~\eqref{eq:spinchain1} and \eqref{eq:spinchain3} as functions of detunings $\varepsilon_L$ and $\varepsilon_R$. (a), (b): $R=150\text{nm}$. (c), (d) $R=44\text{nm}$. (e), (f) $R=35\text{nm}$. Common parameters: $\hbar\omega_0=5 \text{meV}$, $a=22 \text{nm}$.}\label{fig:6}
\end{figure}

Figure~\ref{fig:6} is a pseudo-color plot showing $J_{12}^\text{eff}$ $\left(J_{34}^\text{eff}\right)$, the effective exchange interaction of the left(right) qubit, as functions of both detuning values $\varepsilon_L$ and $\varepsilon_R$.  Figure~\ref{fig:6}(a) and (b) show the case in which $R/a$ is large. We see from Fig.~\ref{fig:6}(a) that $J_{12}^\text{eff}$ is only dependent on $\varepsilon_L$ but not on $\varepsilon_R$, and from Fig.~\ref{fig:6}(b) that $J_{34}^\text{eff}$ only depends on $\varepsilon_R$ but not on $\varepsilon_L$. These results are expected because in this case the two S-T qubits behave independently. On the other hand, the case shown in Fig.~\ref{fig:6}(c) and (d), in which $R/a=2$, reveals very different results. In Fig.~\ref{fig:6}(c), $J_{12}^\text{eff}$ not only depends on $\varepsilon_L$, but also depends weakly on $\varepsilon_R$, leading to a ``bending'' of the stripe showing the increase of the exchange interaction. In particular, as $\varepsilon_R$ is changed from positively to negatively detuned, $\varepsilon_L$ must be detuned further toward the left (negative side), in order to maintain the same level of $J_{12}^\text{eff}$. Similarly, as has been shown in Fig.~\ref{fig:6}(d), while $J_{34}^\text{eff}$ depends mainly on $\varepsilon_R$, it also weakly depends on $\varepsilon_L$: as $\varepsilon_L$ is detuned from negative to the positive direction, one must increase $\varepsilon_R$ a little to maintain the same strength of $J_{34}^\text{eff}$. 
Figure~\ref{fig:6}(e) and (f) show the case in which the two S-T qubits are even closer that  $R/a<2$. In this case, the effect of gate crosstalk that we have already observed becomes more pronounced. Taking Fig.~\ref{fig:6}(e) as an example, as $\varepsilon_R$ is detuned from the positive to negative side, one must detune $\varepsilon_L$ toward the left in order to maintain the exchange interaction for the left S-T qubit. 
We note here that the capacitive coupling between the two qubits does not make significant contribution to the results shown in Fig.~\ref{fig:6} because while the detuning of one qubit is changed from negative to positive values, one accordingly expects a monotonic change of the exchange interaction of the other qubit, and that is not what we have seen in the results.

This gate crosstalk effect, albeit not strong,  is clearly the consequence of the overlap between the electron wave functions in different S-T qubits. Our calculations also show that for $R/a\gg2$, this effect can be safely neglected, but as $R/a$ is reduced close to or less than two, the effect quickly becomes pronounced, as has been shown in Fig.~\ref{fig:6}(c) and (d). Note that $R/a<2$ means that the second dot is even closer to the third dot than the first one, a situation that does not usually happen in experiments. Our results therefore indicates that as long as the S-T qubits are fabricated at a distance which is much greater than the distance between the two dots within one qubit, the ``quantum'' channel of the gate crosstalk is very small and can be safely ignored. However, when the spacing between the adjacent S-T qubits are comparable or smaller than the distance between the two dots within each S-T qubits, one must exercise caution on the gate crosstalk which has appreciable effects on the exchange interactions of the idle qubit.

\section{Conclusions}
\label{sec:conclusion}

In this paper, we have theoretically studied a lateral four-electron four-quantum-dot system constituting two S-T qubits, using molecular-orbital-theoretic methods. We have applied the Hund-Mulliken approximation to keep only the lowest levels in the four dots, but we have explicitly taken into account the admixture of electron wave functions in all dots. We have found that this mixing of wave functions causes interesting consequences on the energy spectrum, exchange interaction and the gate crosstalk of the system. To be specific, we have found that certain states in the energy spectrum are no longer dominated by the computational basis states as expected, when the two S-T qubits are sufficiently close. In this case, the exchange interaction can no longer be straightforwardly found by taking the energy differences between the two relevant levels. On the other hand, we have found that the effective exchange interaction, which is a combinational result of various Hubbard parameters that can be calculated microscopically, agrees well with the exchange interaction calculated by taking energy differences where applicable. This suggests that the exchange interaction in the four-quantum-dot system should be understood in an effective sense. 
We have also quantitatively compared the two commonly conceived schemes coupling the two qubits: the capacitive and the exchange coupling. We found that while one of them shall dominate the two-qubit coupling when the two qubits are either far apart (capacitive coupling) or close-by (exchange coupling), at intermediate distances, found to be $R/a\approx 2.6$ for parameters used in this work, the two kinds of coupling are comparable in strength, complicating the analyses of the evolution of the system. We note here that the value $R/a\approx 2.6$ given here should be considered as an estimate of the critical value given the various levels of approximations used. Therefore when fabricating a sample one should make the inter-qubit distance be considerably smaller or larger than this intermediate regime so that the desired coupling shall dominate.
Last but not least, we have studied the gate crosstalk in the system arising from the overlap between electron wave functions belonging to different qubits. We have found that while this effect is typically very weak, it can become more pronounced when the spacing between the qubits are similar to or even less than the distance between the double dots that constitute the qubit. Our work provides theoretical insights on the manipulation of two S-T qubits on four-quantum-dot devices, which consequently contributes to the effort toward scalable, fault-tolerant quantum computation using spin qubits.

This work was supported by the 
Research Grants Council of the Hong Kong Special Administrative Region, China (No. CityU 21300116) and the National Natural Science Foundation of China (No. 11604277).

\appendix
\setcounter{equation}{0}

\section{Effective Exchange Interaction}\label{Jeff}

In this Appendix, we give a detailed derivation of the effective spin Hamiltonian \eqref{eq:spinchain} from its second-quantized Hubbard-like form.\cite{Yang.11} This has previously been done for a four-quantum-dot system,\cite{Li.12} but our results have subtle differences especially in regard to the occupation-modulated hopping \eqref{eq:fqdH}, which we shall explain below.

The Hamiltonian of a general $n$-electron $n$-dot chain can be written in the second-quantized form as
\begin{equation}
H=H_e+H_t+H_U+H_{J^e}+H_{J^p}+H_{J^t},
\label{eq:fqdHam}
\end{equation}
where
\begin{subequations}
\begin{align}
H_e=&\sum^n_{k=1}\sum_\sigma\epsilon_kc^\dagger_{k\sigma}c^{}_{k\sigma},\label{eq:A6a}\\
H_t=&\sum^{n-1}_{k=1}\sum_\sigma t_{k,k+1}c^\dagger_{k\sigma}c^{}_{k+1,\sigma}+\text{H.c.},\\
H_U=&\sum^3_{k=1}U_{k,k+1}\left(n_{k\uparrow}+n_{k\downarrow}\right)\left(n_{k+1,\uparrow}+n_{k+1,\downarrow}\right)\notag\\
&+\sum^n_{k=1}U_kn_{k\uparrow}n_{k\downarrow},\label{eq:A6c}\\
H_{J^e}=&-\sum^{n-1}_{k=1}\sum_{\sigma_1,\sigma_2}J^e_{k,k+1}c^\dagger_{k\sigma_1}c^\dagger_{k+1,\sigma_2}c^{}_{k+1,\sigma_1}c^{}_{k\sigma_2},\\
H_{J^p}=&-\sum^{n-1}_{k=1}J^p_{k,k+1}c^\dagger_{k+1,\uparrow}c^\dagger_{k+1,\downarrow}c_{k\uparrow}c_{k\downarrow}+ \text{H.c.},\\
H_{J^t}=&-\sum^{n-1}_{k=1}\sum^{k+1}_{i=k}\sum_\sigma J^{t,i}_{k,k+1}n_{i\sigma}c^\dagger_{k\bar{\sigma}}c_{k+1,\bar{\sigma}}+\text{H.c.}.
\label{eq:fqdH}
\end{align}
\end{subequations}

When the Coulomb repulsion is sufficiently large, double occupancy is precluded, and the Hamiltonian can have an effective form as\cite{Li.12}
\begin{equation}
H_\text{eff}=PHP-\sum_Q{PHQ(QHQ-E)^{-1}QHP},
\label{eq:HMeHam}
\end{equation}
where $P=\prod_l(1-n^{}_{l\uparrow}n^{}_{l\downarrow})$ projects the Hilbert space into that of the case in which all dots are singly occupied, $Q=1-P$ is the projection operator for higher energy state, and $E$ is the ground state energy of Hamiltonian. 

Using an intermediate result,
\begin{equation}
\begin{split}
P\sum_{\sigma_1,\sigma_2}c_{l_1\sigma_1}^\dagger c_{l_2\sigma_1}c_{l_2\sigma_2}^\dagger c_{l_1\sigma_2}P=\frac{1}{2}-2\bm{S}_{l_1}\cdot\bm{S}_{l_2},
\label{eq:dqkr}
\end{split}
\end{equation}
elements of Eq.~\eqref{eq:HMeHam} can be evaluated as
\begin{subequations}
\begin{align}
PHP&=\text{Const.}-\sum^n_{k=1}2J^e_{k,k+1}\bm{S}_k\cdot\bm{S}_{k+1},\label{eq:PP}\\
PHQ&=P(H_t+H_{J^t})Q\equiv PH_KQ,
\label{eq:PQ}
\end{align}
\end{subequations}
where
\begin{equation}
\begin{split}
H_K=&\sum^3_{k=1}\sum_{\bar{\sigma}}\left[(t_{k,k+1}-J^{t,k+1}_{k,k+1}n_{k+1,\sigma})c^\dagger_{k\bar{\sigma}}c^{}_{k+1,\bar{\sigma}}\right.\\
&\left.+(t_{k,k+1}-J^{t,k}_{k,k+1}n_{k,\sigma})c^\dagger_{k+1,\bar{\sigma}}c^{}_{k\bar{\sigma}}+\text{H.c.}\right].
\label{eq:PQ1}
\end{split}
\end{equation}

Here, the $J^{t,k+1}_{k,k+1}$ term (and its Hermitian conjugate) refers to the process in which an electron hops from the $(k+1)$th site to the $k$th site and then back, while the $J^{t,k}_{k,k+1}$ term indicates the reverse. In Ref.~\onlinecite{Li.12}, the two processes have not been distinguished. In our work, we have to distinguish the two because as we will see immediately below, the energy change will be very different. The numerator and the denominator of  Eq.~\eqref{eq:HMeHam} are evaluated for the two cases as follows:

(1) For the case with an electron hopping from the $(k+1)$th site to the $k$th and then back, we have
\begin{equation}
\begin{split}
&(PH_KQ\cdot QH_KP)_{k+1\rightarrow k\rightarrow k+1}\\
=&\sum^{n-1}_{k=1}(t_{k,k+1}-J^{t,k+1}_{k,k+1})^2\left(\frac{1}{2}-2\bm{S}_k\cdot\bm{S}_{k+1}\right),
\label{eq:PQQP3}
\end{split}
\end{equation}
and
\begin{equation}
\begin{split}
(QHQ-E)_{k+1\rightarrow k\rightarrow k+1}=&\text{}U_{k}+U_{k-1,k}-U_{k,k+1}\\
&-U_{k+1,k+2}+\epsilon_k-\epsilon_{k+1}.
\label{eq:PQQP4}
\end{split}
\end{equation}

 (2) For the opposite case in which an electron transits from the $k$th site to the  $(k+1)$th and then back, we have
\begin{equation}
\begin{split}
&(PH_KQ\cdot QH_KP)_{k\rightarrow k+1\rightarrow k}\\
=&\sum^{n-1}_{k=1}(t_{k,k+1}-J^{t,k}_{k,k+1})^2\left(\frac{1}{2}-2\bm{S}_k\cdot\bm{S}_{k+1}\right),
\label{eq:PQQP1}
\end{split}
\end{equation}
and
\begin{equation}
\begin{split}
(QHQ-E)_{k\rightarrow k+1\rightarrow k}=&\text{}U_{k+1}+U_{k+1,k+2}-U_{k-1,k}\\
&-U_{k,k+1}+\epsilon_{k+1}-\epsilon_k.
\label{eq:PQQP2}
\end{split}
\end{equation}

Summarizing the results, we have transformed the Hamiltonian \eqref{eq:fqdHam} into the Heisenberg spin model, defined as
\begin{equation}
\begin{split}
H_{\text{eff}}&=\sum^n_{k=1}J_{k,k+1}\bm{S}_k\cdot\bm{S}_{k+1},
\label{eq:HH}
\end{split}
\end{equation}
with the exchange interaction being
\begin{equation}
\begin{split}
J_{k,k+1}=&\frac{2(t_{k,k+1}-J^{t,k}_{k,k+1})^2}{U_{k}+U_{k-1,k}-U_{k,k+1}-U_{k+1,k+2}+\epsilon_k-\epsilon_{k+1}}\\
&+\frac{2(t_{k,k+1}-J^{t,k+1}_{k,k+1})^2}{U_{k+1}+U_{k+1,k+2}-U_{k-1,k}-U_{k,k+1}+\epsilon_{k+1}-\epsilon_k}\\
&-2J^e_{k,k+1}.
\label{eq:Jeff1}
\end{split}
\end{equation}

As a last remark, we note that whenever one encounters an out-of-range index in the parameters of Eq.~\eqref{eq:Jeff1}, that parameter should be treated as zero, e.g. $U_{n,n+1}=U_{01}=0$.


%

\end{document}